\documentclass[preprint,12pt]{elsarticle}

\usepackage{graphicx}
\usepackage{dcolumn}
\usepackage{bm}
\usepackage{amssymb}

\def\be{\begin{equation}}
\def\ee{\end{equation}}
\def\e#1{\label{#1}\end{equation}}
\def\bea{\begin{eqnarray}}
\def\eea{\end{eqnarray}}
\def\ea#1{\label{#1}\end{eqnarray}}

\def\bem#1{\begin{mathletters}\label{#1}}
\def\eml{\end{mathletters}}

\def\ket#1{{|#1\rangle}}

\def\4#1{{\boldsymbol{#1}}}
\def\8#1{{\widetilde{#1}}}
\def\bse{\begin{subequations}}
\def\ese{\end{subequations}}
\def\eqref#1{{(\ref{#1})}}

\journal{Optics Communications}

\begin{document}

\begin{frontmatter}

\title{Quantum cryptography using partially entangled states}

\author{Goren Gordon$^a$ and Gustavo Rigolin$^{b,}$\fnref{1}}
\address{$^a$ Department of Chemical Physics, Weizmann Institute of Science, Rehovot 76100,
Israel} \address{$^b$ Departamento de Fisica, Universidade Federal
de Sao Carlos, Caixa Postal 676, Sao Carlos, 13565-905, SP,
Brazil}\fntext[1]{rigolin@ufscar.br}


\begin{abstract}
We show that non-maximally entangled states can be used to build a
quantum key distribution (QKD) scheme where the key is
probabilistically teleported from Alice to Bob. This probabilistic
aspect of the protocol ensures the security of the key without the
need of non-orthogonal states to encode it, in contrast to other
QKD schemes.  Also, the security and key transmission rate of the
present protocol is nearly equivalent to those of standard QKD
schemes and these aspects can be controlled by properly harnessing
the new free parameter in the present proposal, namely, the degree
of partial entanglement. Furthermore, we discuss how to build a
controlled QKD scheme, also based on partially entangled states,
where a third party can decide whether or not Alice and Bob are
allowed to share a key.
\end{abstract}

\begin{keyword}
Quantum communication \sep  Quantum cryptography and communication
security \sep Entanglement production and manipulation \PACS
03.67.Hk \sep 03.67.Dd \sep 03.67.Bg
\end{keyword}

\end{frontmatter}


\section{Introduction}

The ability to communicate secretly is considered one of the most
important challenges of the information era \cite{codebook}. For
all practical purposes most modern public key systems can be
considered secure \cite{publickey}. However, this security is not
based on any mathematical proof but on the belief that there is no
classical algorithm to factorize huge prime numbers in a
reasonable amount of time. Some classical cyphers, such as the
one-time pad \cite{codebook}, do not have the aforementioned
problem. They are private key protocols whose security is entirely
based on a random string of bits, the key, which only the sender
(Alice) and the receiver (Bob) should know. Once the key is
secretly transmitted the communication is absolutely secure. The
drawback with these cyphers is that any key transmitted through a
classical channel can be passively monitored. Although it may be
technologically difficult to get the key without being noticed, it
is in principle possible.

The solution of the key transmission problem based on the laws of
physics was presented for the first time by Bennett and Brassard
in a seminal work \cite{bb84}. Making use of quantum channels
(polarized photons) the authors theoretically showed the
possibility to share a secret key with absolute security if the
laws of quantum mechanics are correct. They have shown that any
interference of an eavesdropper (Eve) on the quantum channel can
be detected by Alice and Bob at the end of the protocol. Other
interesting schemes were later proposed \cite{ekert,bbm92}, in
particular the E91 protocol which was the first QKD scheme that
employed maximally entangled states \cite{ekert}. For an extensive
review on other protocols and on experimental feasibilities we
refer the reader to Ref. \cite{Gis01}.

In this contribution we present a new QKD scheme that uses
directly non-maximally entangled states (no entanglement
concentration needed) and the probabilistic quantum teleportation
protocol (PQT) \cite{aga02,gor06c}. Nevertheless, the present
scheme resembles the BB84 \cite{bb84} rather than the E91
\cite{ekert} protocol, i.e., although we make use of entanglement
there is no need to check for violation of any Bell inequality to
assure the security of the shared key. It is the probabilistic
aspect of the PQT that guarantees the security of the teleported
key and, as we show later, also allows it to be encoded in a set
of orthogonal states. Note that in BB84-like protocols it is
mandatory to encode the key in non-orthogonal states to make sure
its transmission is secure.

In contrast to other protocols, where departure from maximal
entanglement makes them inoperable, our scheme exploits partial
entanglement and, using a special generalized Bell measurement,
ensures flawless key distribution. Furthermore, this new QKD
scheme takes advantage of a new free parameter, namely the degree
of partial entanglement, that enables more control over the
security and transmission rate of the protocol. Indeed, as
explained later, this freedom allows us to introduce a minor
modification in the QKD scheme that turns it into a controlled QKD
protocol, where a third party (Charlie) has the final word on
whether or not Alice and Bob are allowed to share a secret key,
even after all steps of the protocol were implemented.  It is also
worth mentioning that Charlie decides whether or not Alice and Bob
will share a key without ever knowing it, a feature that has
practical applications. We also show other possible interesting
extensions of the basic protocol and how we can improve its
security and the reliability of the transmitted key.

\section{The tools}

One important ingredient in this QKD protocol, and the one that
allows it to depart from E91-like protocols, is the use of
\textit{partially} entangled states to transmit the secret key
from one party to the other. Indeed, as we will show, by playing
with different kinds of partially entangled states and with
different joint measurement basis, Alice can teleport to Bob a
secret key. The other ingredient is, as anticipated in the last
sentence, the proper use of a probabilistic teleportation
protocol, which allows us to harness the teleporting power of a
non-maximally pure entangled state.

Let us start by recalling the PQT as developed in Ref.
\cite{aga02} and extended in Ref. \cite{gor06c}. As usual, Alice
wants to teleport the following qubit to Bob,
\begin{equation}
\ket{\phi^A} = \alpha \ket{0} + \beta \ket{1},
\end{equation}
where $\alpha$ and $\beta$ are arbitrary complex numbers such that
$|\alpha|^2$ + $|\beta|^2$ $=$ $1$. Contrary to the original
proposal \cite{Ben93} Alice and Bob now share a non-maximally
entangled state,
$ \ket{\Phi_{n}^{+}} = N(\ket{00} + n\ket{11}), $
with $0<|n|<1$ and $N=1/\sqrt{1+|n|^2}$, which naturally leads to
the following orthonormal basis,
\bea
\ket{\Phi_{m}^{+}} & = &(\ket{00} + m \ket{11})/\sqrt{1+|m|^2},
\label{phiplus}\\
\ket{\Phi_{m}^{-}} & = & (m^{*}\ket{00} - \ket{11})/\sqrt{1+|m|^2},\\
\ket{\Psi_{m}^{+}} & = & (\ket{01} + m \ket{10})/\sqrt{1+|m|^2},\\
\ket{\Psi_{m}^{-}} & = & (m^{*}\ket{01} - \ket{10})/\sqrt{1+|m|^2},
\label{psiminus}
\eea
%
with $m^*$
denoting the complex conjugate of $m$ and $M=1/\sqrt{1+|m|^2}$.
Using the generalized Bell states (GBS) above we can rewrite the
three qubit state belonging to Alice and Bob as,
%
%
%
$\ket{\Phi}$ $=$ $\ket{\phi^A}$ $\otimes$ $\ket{\Phi_{n}^{+}}$
 $=$  $MN$ $(\ket{\Phi^+_m}$ $(\alpha\ket{0}$ $+$ $nm^{*}\beta\ket{1})$
$+$ $\ket{\Phi^-_m}$ $(m\alpha\ket{0}$ $-$ $n\beta\ket{1})$ $+$
$\ket{\Psi^+_m}$ $(m^{*}\beta\ket{0}$ $+$ $n\alpha\ket{1})$ $+$
$\ket{\Psi^-_m}$ $(-\beta\ket{0}$ $+$ $nm\alpha\ket{1}))$,
with the first two qubits being with Alice and the last one with
Bob. Alice now proceeds by implementing a generalized Bell
measurement (GBM), i.e., she projects her two qubits onto one of
the four GBS. 
(Ref. \cite{Kim04} discusses three possible ways to experimentally
implement a GBM.) The probability to obtain a given GBS is,
\bea
P_{\Phi^+_m}&=&\left(|\alpha|^2+|mn\beta|^2\right)/[(1+|m|^2)(1+|n|^2)], \\
P_{\Phi^-_m}&=&\left(|m\alpha|^2+|n\beta|^2\right)/[(1+|m|^2)(1+|n|^2)], \\
P_{\Psi^+_m}&=&\left(|n\alpha|^2+|m\beta|^2\right)/[(1+|m|^2)(1+|n|^2)], \\
P_{\Psi^-_m}&=&\left(|mn\alpha|^2+|\beta|^2\right)/[(1+|m|^2)(1+|n|^2)].
\eea

Alice then sends Bob the result of her measurement (two bits) via
a classical channel who, whereupon, applies a unitary
transformation on his qubit according to this information. These
transformations are the same ones given in the original
teleportation protocol \cite{Ben93}: If Alice gets
$\ket{\Phi^{+}_m}$ then Bob does nothing, if she gets
$\ket{\Phi^{-}_m}$ he applies a $\sigma_{z}$ operation, if Alice
measures $\ket{\Psi^{+}_m}$ then Bob applies $\sigma_{x}$ and
finally for $\ket{\Psi^{-}_m}$ he applies $\sigma_{z}\sigma_{x}$.
Here, $\sigma_z$ and $\sigma_x$ are the usual Pauli matrices
($\sigma_z|0(1)\rangle$ $=$ $+(-)|0(1)\rangle$ and
$\sigma_x|0(1)\rangle$ $=$ $|1(0)\rangle$).
After the correct transformation Bob's qubit is given by one of
the following possibilities,
\bea \ket{\Phi^{+}_m} &
\longrightarrow & \ket{\phi^B}=
\frac{\alpha\ket{0}+ nm^{*}\beta\ket{1}}{\sqrt{|\alpha|^2+|mn\beta|^2}},
\label{A} \\
\ket{\Phi^{-}_m} & \longrightarrow & \ket{\phi^B}=
\frac{m\alpha\ket{0}+ n\beta\ket{1}}{\sqrt{|m\alpha|^2+|n\beta|^2}},
\label{B} \\
\ket{\Psi^{+}_m} & \longrightarrow & \ket{\phi^B}=
\frac{n\alpha\ket{0}+m^{*}\beta\ket{1}}{\sqrt{|n\alpha|^2+|m\beta|^2}},
\label{C} \\
\ket{\Psi^{-}_m} & \longrightarrow & \ket{\phi^B}=
\frac{mn\alpha\ket{0}+\beta\ket{1}}{\sqrt{|mn\alpha|^2+|\beta|^2}}.
\label{D}
\eea
From now on, and without loss of generality \cite{gor06c}, we
consider $n$ and $m$ to be real quantities. Therefore, looking at
Eqs.~(\ref{B}) and (\ref{C}) we realize that if $n=m$ the state
with Bob is $\alpha \ket{0} + \beta \ket{1}$ and the protocol
works perfectly. There exist other possibilities, which come
from Eqs.~(\ref{A}) and (\ref{D}), namely, $nm=1$ or $nm^*=1$. But
this is only possible if we have maximally entangled states since
those relations imply $|n|=|m|=1$.
For $n=m$ the probability of success is simply
\be
P_{suc}=P_{\Phi^-_n}+P_{\Psi^+_n}=\frac{2n^2}{(1+n^2)^2}.
\label{ProbSuc}
\ee
Therefore, if Alice knows the entanglement of the channel, which
increases monotonically with $n$, she can match her measurement
basis ($m=n$) in order to make the protocol work with probability
$P_{suc}$. It is important to note that if $n\neq m$ Bob obtains a
different state and the protocol fails. It is this property that
we explore in order to build our QKD scheme.

\section{The QKD scheme}

\begin{figure}[!ht]
\centering\includegraphics[angle=-90,width=8.5cm]{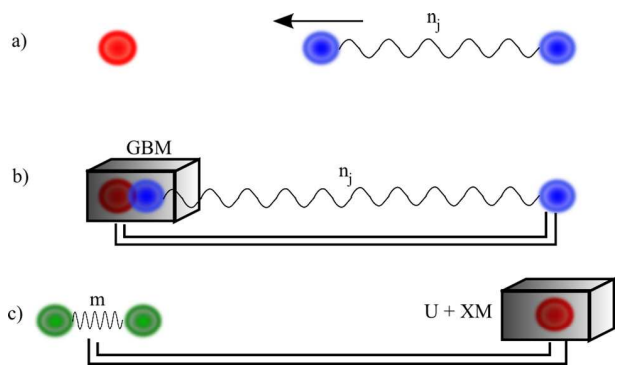}
\caption{(Color online) Schematic representation of one run of the
QKD protocol. a) Bob prepares randomly a partially entangled
state, either $\ket{\Phi^+_{n_1}}$ or $\ket{\Phi^+_{n_2}}$,
without telling Alice which one. Then he sends her one of the
entangled qubits. b) Alice implements the PQT choosing randomly
between two generalized Bell measurements (GBM). She teleports
randomly either the state $\ket{+}$ or $\ket{-}$. She then tells
Bob her measurement result, $|\Phi_m^{\pm}\rangle$ or
$|\Psi_m^{\pm}\rangle$, but does not tell him which basis she used
(if $m=n_1$ or $m=n_2$). c) With this information Bob applies the
right unitary operation (U) on his qubit and projects it onto the
X-basis $|\pm\rangle$ (XM). Then both parties broadcast the values
of $n$ and $m$. If $n\neq m$ they discard this run. If $n=m$ and
Alice's GBM yielded $|\Phi^-\rangle$ or $\ket{\Psi^+}$ they both
agree on the teleported qubit, which constitute one bit of the
secret key.}\protect\label{Fig1}
\end{figure}

Let us assume that Bob prepares with equal probability two
partially entangled states, $\ket{\Phi_{n_1}^{+}}$ $=$
$N_1(\ket{00} + n_1\ket{11})$  and $\ket{\Phi_{n_2}^{+}}$ $=$
$N_2(\ket{00} + n_2\ket{11})$, where $n_1\neq n_2$ and
$N_j=1/\sqrt{1+n_j^2}$, $j=1,2$. For each state he keeps one qubit
and send the other one to Alice (See Fig. \ref{Fig1}).
Both parties previously agreed on the values of $n_1$ and $n_2$
but at this stage Bob does not tell Alice the respective value of
$n$ for each entangled state he prepared. Alice, on the other
hand, prepares randomly two types of single qubit states,
$\ket{\pm}$ $=$ $(\ket{0} \pm \ket{1})/\sqrt{2}$, which are to be
associated with the secret key she wants to share with Bob.  For
example, the parties use the convention that $\ket{+}$ represents
the bit $0$ and that $\ket{-}$ the bit $1$. Note that we do not
need another encoding for the bits $0$ and $1$ that is
non-orthogonal to the previous one, as required by the BB84
protocol \footnote{We could have chosen another encoding for the
bits $0$ and $1$. This would increase the security of the present
protocol on the expense of the key transmission rate. However, as
it is clear from the security analysis, this is not mandatory.}.
Alice then uses each qubit received from Bob to implement the PQT
for each one of her randomly generated states $\ket{\pm}$. In
doing so, she also chooses in a random way whether to project each
pair of qubits (hers and Bob's) to the GBS with $m=n_1$ or
$m=n_2$. Alice, however, does not inform Bob of the value of $m$
but only which GBS she gets.
At the end of this
stage Bob knows what her measurement results were (but not $m$),
which allows him to implement the right unitary operation on his
qubits. After that, each one of his qubits are described by one of
the four states given by Eqs.~(\ref{A})-(\ref{D}), with
$\alpha=\beta=1/2$.

The last six steps of the protocol are as follows. First, Bob
projects his qubits onto the $\ket{\pm}$ basis. Second, Bob and
Alice reveal in a public channel the following information. Bob
tells Alice which value of $n$ ($n_1$ or $n_2$) he has assigned to
each partially entangled state whilst Alice tells him the value of
$m$ ($n_1$ or $n_2$) for each GBM she made. Third, they keep all
the cases where she has rightly matched the entanglement of the
channel with the entanglement of the measuring basis, i.e.,
whenever the shared entangled state was $\ket{\Phi_{n_j}^{+}}$ and
Alice chose $m=n_j$, $j=1,2$. Fourth, they discard all the other
cases since the PQT fails there ($m\neq n_j$). Fifth, within the
cases where the matching condition is satisfied, Bob and Alice
keep only those instances where her measurement results were
$\ket{\Phi^{-}}$ or $\ket{\Psi^{+}}$, the so called successful
runs of the PQT.
For those, and only those runs of the protocol, Alice and Bob are
sure they agree on the teleported state and consequently on the
random string of zeros and ones. Finally, the last step consists
of using half of the successful cases to test whether or not Eve
has tried to tamper with the key.

In an idealized situation, i.e., perfect detectors and no noise,
they simply broadcast half of their valid results in a public
classical channel and check if they always get the same bits. If
they do, the remaining half of bits are their secret key. If they
fail to agree on the public data, they discard everything and
repeat the whole protocol again. However, noise and non-ideal
detectors will introduce some errors even when all the steps of
the protocol are successful. Nevertheless, Alice and Bob can still
achieve any desired level of security by increasing the size of
the shared key and employing classical reconciliation protocols
and privacy amplification techniques already developed for other
QKD schemes \cite{Ben92}.

Assuming an ideal scenario, for instance, excellent detectors and
efficient measurement processes, we can calculate the maximum rate
of how many teleported qubits constitute the final secret key. We
know that Alice implements the PQT half of the times making a GBM
with $m=n_1$ and half with $m=n_2$. Therefore, the total
probability of success for all PQT is, according to
Eq.~(\ref{ProbSuc}),
$n_1^2/(1+n_1^2)^2 + n_2^2/(1+n_2^2)^2.$ But half of the
successful cases are discarded to check for the presence of Eve,
and the final rate becomes
\begin{equation}
P_{suc}^{final} =\frac{n_1^2}{2(1+n_1^2)^2} +
\frac{n_2^2}{2(1+n_2^2)^2}. \label{finalrate}
\end{equation}
On the other hand, if we look at the BB84 protocol \cite{bb84}, we
see that half of the times Bob measures the qubits he receives
from Alice using the same basis she employed to prepare them and,
within these successful runs, the other half is used to test for
the security of the protocol. This gives us, assuming no loss
during the transmission of the qubit and ideal detectors, a total
idealized rate of $1/4$. Returning to the protocol presented here,
it is not difficult to see that $P_{suc}^{final}<1/4$, no matter
what the values of $n_1$ and $n_2$ are. (If they are equal to one
we have $1/4$ but then the protocol is useless.) However, for
modest values of $n_1$ and $n_2$ (a little greater than $0.5$) we
get rates above $15\%$. If we allow one of them to approach unity
we do even better. For example, if we have $n_1=0.5$ and $n_2=0.9$
we already obtain rates higher than $20\%$.

There exists, nevertheless, an important feature that we can
easily achieve employing this protocol that is unattainable using
the BB84 protocol. We can transform it into a sort of controlled
QKD scheme introducing another party (Charlie) who can decide
whether or not Alice and Bob are allowed to share a secret key
even after they finished all steps of the protocol. In order to do
that, we let Charlie prepare and distribute the entangled states
$|\Phi_{n_1}^{+}\rangle$ and $|\Phi_{n_2}^+\rangle$ to Alice and
Bob. Hence, if Charlie publicly announces the values of $n$ for
each entangled state he prepared he can make the protocol work
without ever knowing the key. Otherwise, if he does not broadcast
this information, the protocol ultimately fails. Note that the
probability of success in this scenario, assuming Charlie
broadcast all the values of $n$ for each entangled pair he
prepares, is the same we had before, Eq.~(\ref{finalrate}). His
role here is simply to distribute the entangled states between
Alice and Bob, without changing the final success rate for the
protocol. We also remark that a similar third party control can be
achieved using a different QKD scheme based on maximally entangled
Bell states \cite{Dur08}.

At this point we wish to emphasize the main differences between
the present scheme and the BB84 and the E91 protocol. As described
above, here we can achieve a level of third party control that is
unattainable using the former two protocols. This is an important
and practical characteristic of this scheme that, as we show
below, can also be extended to a fourth, fifth, \ldots, $n$-th
party level of control. Moreover, in the present protocol the key
is never transmitted from Alice to Bob as in the BB84 protocol.
Rather, it is teleported from one party to the other, which gives
an additional flexibility for this protocol in its third party
formulation. Indeed, once Alice and Bob have shared the partially
entangled states distributed by Charlie they can easily exchange
their roles. Instead of Alice teleporting the key to Bob, he is
the one who teleports the key to her. Also, contrary to the E91
protocol where a maximally entangled state is directly responsible
to the generation of the secret key, here we use a non-maximally
entangled state as a channel through which the key is teleported.
In other words, the non-maximally entangled states of the present
scheme have no direct role on the generation of the secret key.

\section{Security}

The security of this protocol is based on the same premises of the
BB84 protocol and, therefore, we can understand the security of
the former by recalling the security analysis \cite{bb84} of the
latter. The key ingredient here is the recognition that there are
two unknown sets of actions throughout the implementation of the
BB84 protocol that are only publicly revealed at the end of it:
the basis in which Alice prepared her qubits and the basis in
which Bob measured the qubits received from Alice. A similar thing
happens for the present protocol. We have two unknown sets of
actions throughout each run of the protocol that are revealed only
at its end: the entanglement of the shared qubits between Alice
and Bob and what basis Alice used to implement the GBM. This lack
of information prevents Eve from always obtaining the right bit
being sent from Alice to Bob without being noticed. As we show
below, the laws of quantum mechanics forbid Eve from acquiring
information about the key being transmitted without disturbing the
quantum state carrying it if she does not know which entangled
state is shared between Alice and Bob.

Let us assume, for definiteness, that in one of the runs of the
protocol Alice prepared the state $|+\rangle$ and that Eve,
somehow, replaced the entangled state produced by Bob with one
produced by her. Eve wants the state prepared by Alice to be
teleported to her. By doing so she thinks she can obtain
information on the key. However, she does not know which GBM Alice
implemented to perform the PQT ($m=n_1$ or $m=n_2$). This
information is only revealed after Bob confirms he measured his
qubit. She only knows that the measurement result of Alice is,
say,  $|\Phi^-\rangle$. (This is the best scenario for Eve.)
Therefore, Eve's qubit is described by Eq.~(\ref{B}),
$\ket{\phi^E}= \frac{m\ket{0}+ n\ket{1}}{\sqrt{m^2+n^2}},$ which
can be written as,
\begin{equation}
\ket{\phi^E} = \frac{1}{\sqrt{2(m^2+n^2)}}\left[ (m+n)\ket{+} +
(m-n)\ket{-}\right]. \label{stateEve}
\end{equation}
Looking at Eq.~(\ref{stateEve}) we see that unless Eve guessed
correctly Alice's choice for $m$ (and this only happens half of
the times), preparing the right entangled state
 with $n=m$, we have a superposition
of the states $\ket{+}$ and $\ket{-}$. This implies that she
obtains the wrong bit being transmitted with probability
$P_{wrong}=(m-n)^2/(2(m^2+n^2))$ $=$ $1/2 - mn/(m^2+n^2)$. In
other words, since we have a superposition of the right and wrong
answers quantum mechanics forbids Eve from always getting the
right one with a single measurement. It is clear now that this is
similar to the argument used to prove the security of the BB84
protocol. Hence, no matter what Eve does, if she prepared the
wrong entangled state and Bob the right one, she will be caught
trying to tamper with the key when Alice and Bob publicly compare
part of it. This is true since Eve cannot with certainty send Bob
another qubit which mimics the right one. Furthermore,
Eq.~(\ref{stateEve}) tells us that the greater the difference
between $n_1$ and $n_2$ the more likely will Eve be detected. We
can see this by noting that $P_{wrong}$ increases as a function of
$|m-n|$ or, equivalently, as a function of the difference in
entanglement between the channels. Lastly, the chances of Eve
being caught also increases with the size of the string of bits
being publicly announced.

We can also estimate the optimal range of parameters ($n_1$ and
$n_2$) for this protocol, assuming we want to maximize the
transmission rate while at the same time minimizing the chances of
Eve guessing the correct qubit being teleported. In other words,
we want to maximize a function proportional to $P_{wrong}P_{suc}$,
where $P_{wrong}$ is, as given above, the probability of Eve
guessing the wrong qubit and $P_{suc}$ is Eq.~(\ref{finalrate}),
the total rate of success in the transmission of the key. Both
$P_{wrong}$ and $P_{suc}$ are now considered functions of $n_1$
and $n_2$. A simple numerical analysis shows that the best
strategies occur for $n_1 \approx 1$ and $n_2\approx 0$ (and
vice-versa), while the worst cases for Alice and Bob occur when
$n_1 \approx n_2$.

Note that the security check outlined above is an idealization. In
real-life situations we always have noise and imperfect devices
that give wrong answers even in the absence of Eve. However, this
can be controlled using classical reconciliation protocols and
privacy amplification \cite{Ben92}. A more detailed security analysis
based on bounds for the mutual information between Alice, Bob, and
Eve using, for example, the techniques of Refs. \cite{Bru00,Kar05}, is beyond our
goals here and is left for future work.

\section{Extensions of the QKD scheme}

The QKD scheme presented here is very versatile and allows for
arbitrary control over the protocol parameters. This can be
achieved by introducing two extensions, where one increases its
security and the other increases the distance of reliable
transmission of the key.
The security of the protocol is increased by allowing Bob to
generate more than two partially entangled states. For example,
instead of just creating the states $\ket{\Phi_{n_j}^+}$, $j=1,2$,
he can create three or more states with different $n$. With only
two states, Eve can guess the right GBM in half of the successful
runs of the PQT. However, with more entangled states, her chances
are reduced to $1/N$, where $N$ is the number of partially
entangled states produced by Bob. On the other hand, this increase
in security reduces the transmission rate of the key since it
becomes less likely that Alice and Bob achieve the matching
condition ($m=n$).

To extend the distance of reliable key transmissions we can use
quantum repeater \cite{Bri98} stations. In this scenario it is the
first station (the closest to Alice) that generates the partially
entangled states and then publicize the values of $n$, only after
Bob measures his qubits. The other stations use maximally
entangled states to successively teleport Alice's qubit to Bob.
Note that security increases if other repeater stations use
partially entangled states too.
The repeater stations can also be used to extend the third party
control described before to any number of parties. Indeed, if we
allow each station to freely choose its own partially entangled
states we are increasing the number of parties that can decide
whether or not Alice and Bob will share a secret key. This is true
for the protocol will work if, and only if, all the repeater
stations disclose to Alice and Bob which partially entangled
states they generated at each run of the protocol.

\section{Experimental feasibility}

While noise and decoherence of entangled qubits usually result in
mixed states \cite{yu04,gor08EPL}, partially-entangled states used
in the aforementioned QKD protocol can be considered in the
scenario of coupling to a zero-temperature bath \cite{gor06b}. In
this regime dynamical control of decoherence
\cite{kof00,gor08PRL,gor06b} allows one to determine the amount of
partial entanglement of the channel by properly tuning
the relative decoherence between the qubits. Thus, the party
sending the partially entangled states (either Bob in the standard
QKD or Charlie in the controlled version) can select
the degree of partial entanglement and is not restricted by
the amount of noise in the system.

\section{Summary}

We showed that partially entangled states are useful resources for
the construction of a direct QKD scheme by the proper use of
probabilistic teleportation protocols.
This has an interesting implication on the practical
implementation of entangled based QKD schemes, as it is extremely
difficult to produce maximally entangled qubits. Using the
protocol presented here, one can alleviate the experimental
demands on the production of entangled pairs without rendering QKD
inoperable. Furthermore, the present partially entangled state
based QKD scheme is flexible enough that we were able extend it in
at least three directions, each one augmenting its usability.
The first one
turned the protocol into a controlled QKD scheme, where a third
party decides whether or not Alice and Bob are able to share a
secret key. Then we demonstrated how one can increase its security by
letting the parties use more and more different partially
entangled states to implement one of the steps of the QKD
protocol, namely, the probabilistic teleportation protocol. And
third, we discussed how the use of quantum repeaters extends the
distance of reliable key transmission without diminishing the key
rate.

Finally, the present QKD protocol naturally leads to new
interesting questions. For instance, can we increase the key
transmission rate using partially entangled qudits instead of
qubits? Is there any possible way of devising a similar approach
using mixed entangled states? Or using continuous variable
entangled systems? Can we do better by using different types of
entangled qubit-like states, such as the cluster state
\cite{Bri01} or the cluster-type coherent entangled states
\cite{Mun08}? It is our hope that the ideas presented here might
lead to clues on how to answer these questions.

\section*{Acknowledgments}
G. R. thanks the Brazilian agency Coordena\c{c}\~ao de
Aperfei\c{c}oamento de Pessoal de N\'{\i}vel Superior (CAPES) for
funding this research.


\end{document}